\documentclass[11pt,fleqn]{article}
\usepackage{amsmath,graphicx}

\usepackage{amsfonts,amssymb,cite}
\usepackage{epsf,graphicx}

\def\noi{\noindent}

\newcommand{\Title}[1]{\noi {{\Large\bf #1}}\\[1ex]}

\def\Aunames#1{\noi{\bf #1}}
\def\auth#1{${}^{#1}$}
\def\Addresses#1{\medskip\noi \protect
	\begin{description}\itemsep -3pt {\it #1} \end{description}}
\def\addr#1#2{\item[${}^{#1}$]{\it #2}}

\def\email#1#2{\footnotetext[#1]{e-mail: #2}\addtocounter{footnote}{1}}

\def\nqq{\hspace*{-2em}}

\def\Jl#1#2{#1 {\bf #2},\ }

\def\ApJ#1 {\Jl{Astroph. J.}{#1}}
\def\CQG#1 {\Jl{Class. Quantum Grav.}{#1}}
\def\DAN#1 {\Jl{Dokl. AN SSSR}{#1}}
\def\GC#1 {\Jl{Grav. Cosmol.}{#1}}
\def\GRG#1 {\Jl{Gen. Rel. Grav.}{#1}}
\def\JETF#1 {\Jl{Zh. Eksp. Teor. Fiz.}{#1}}
\def\JETP#1 {\Jl{Sov. Phys. JETP}{#1}}
\def\JHEP#1 {\Jl{JHEP}{#1}}
\def\JMP#1 {\Jl{J. Math. Phys.}{#1}}
\def\NPB#1 {\Jl{Nucl. Phys. B}{#1}}
\def\NP#1 {\Jl{Nucl. Phys.}{#1}}
\def\PLA#1 {\Jl{Phys. Lett. A}{#1}}
\def\PLB#1 {\Jl{Phys. Lett. B}{#1}}
\def\PRD#1 {\Jl{Phys. Rev. D}{#1}}
\def\PRL#1 {\Jl{Phys. Rev. Lett.}{#1}}

\def\lal{&&\nqq {}}

\def\beq{\begin{equation}}
\def\eeq{\end{equation}}
\def\bear{\begin{eqnarray}}
\def\bearr{\begin{eqnarray} \lal}
\def\ear{\end{eqnarray}}
\def\earn{\nonumber \end{eqnarray}}

\begin{document}

\Title{Extended geometry and kinematics 
induced by \\ biquaternionic and twistor structures}

\Aunames{Vladimir V. Kassandrov\auth{a,1} and Nina V. Markova\auth{b,2}}

\Addresses{
\addr a {Institute of Gravitation and Cosmology, Peoples' Friendship
	University of Russia, Moscow, Russia}
\addr b {Department of Applied Mathematics, Peoples' Friendship
	University of Russia, Moscow, Russia}
	}

\begin{abstract}
  The algebra of biquaternions possess a manifestly Lorentz invariant form and induces an extended space-time geometry. We consider the links between this complex pre-geometry and real geometry of  the Minkowski space-time. Twistor structures naturally arise in the framework of biquaternionic analysis. Both together, algebraic and twistor structures impose rigid restriction on the transport of singular points of  biquaternion-valued fields identified with particle-like formations. 
\end{abstract}
\section{Biquaternions as the algebra of extended  space-time}
\label{sec1}
The algebra of complex quaternions, {\it biquaternions}, $\mathbb B$ is the most appropriate candidate for the role of a {\it space-time algebra} (the notion introduced firstly by D. Hestenes in~\cite{Hestenes}). Indeed, its symmetry (automorphism) group $SO(3,\mathbb C)$ is isomorphic to the spinor Lorentz group $SL(2,\mathbb C)$, and its multiplication table can be represented in a manifestly Lorentz invariant form~\cite{Grgin, RelAlg16}. However, vector space of $\mathbb B$ has four {\it complex} dimensions  and does not correspond therefore to the structure of real Minkowski space {\bf M}. Physical sense of four ``superfluous'' coordinates (if any) remains indefinite, and we shall concentrate on this problem below. 

Nonetheless, biquaternions and $\mathbb B$-valued functions, as well as the structure of extended complex space-time $\mathbb C\bf M$ itself, have been exploited in a number of interesting approaches in the field theory and general relativity. Particularly, in GTR complex space-time serves as a source of null shear-free congruences (NSFC) on its ``real cut'' which are known to determine the structure of Robinson-Trautman metrics. Moreover, consideration of a pointlike source ``moving'' in a complex extension $\mathbb C\bf M$ results in a number of interesting conclusions about its characteristics and characteristics of particlelike formations -- caustics of corresponding NSFC arising on the real cut~\cite{Newman1,Newman2,Burin1}. A well-known example of this situation is the anomalous Dirac value~\cite{Carter} of the  gyromagnetic ratio of the {\it Kerr singular ring}, the caustic of corresponding NSFC.  The latter property has been used in a number of {\it models of electron}, in particular the ``Dirac -- Kerr-Newman electron'' elaborated by 
A. Burinskii~\cite{Burin2}. 

As for the biquaternions themselves, there exists a wide variety of works devoted to various applications of them in field theory and physical geometry (for a review see, e.g., ~\cite{Gsponer1, Gsponer2}). In particular, the quaternionic version of Special Relativity, based on the properties of $\mathbb B$ and the concept of 3-dimensional time, has been elaborated in the works of A. Yefremov~\cite{Yefrem1,Yefrem2}. There is also a lot of attempts (see \cite{Gsponer1, Gsponer2}) in which physical fields are regarded as $\mathbb B$-valued functions subject to a kind of {\it differentiablity conditions}. The latter, usually constructed in a full analogy with the Cauchy-Riemann (CR) conditions  from complex analysis, play the role of the primary field equations. 
 
 However, all these generalizations reproduce the linear CR-structure, do not take in account the {\it non-commutativity} of $\mathbb B$ and, moreover, are not mathematically justified (value of a derivative usually depends on the direction of infinitesimal increment, contrary to the complex case). 
 
On the other hand, in our version of non-commutative analysis (see, e.g., ~\cite{GR95, Qanalisys, YadPhys} and references therein) non-linearity of the differentiability conditions for (bi)quaternionic functions follow directly from the property of $\mathbb B$ non-commutativity, so that $\mathbb B$-differentiable functions can be actually treated as (self-)interacting physical fields.

Moreover, the system of $\mathbb B$-differentiability equations possess natural 2-spinor and twistor structures, and its general solution represents in fact an invariant version of the celebrated Kerr-Penrose theorem~\cite{Kerr, Penrose}. The latter gives a complete algebraic description  of the NSFC on the  Minkowski $\bf M$ or Kerr-Shild spaces~\cite{Kerr} via twistor generating functions. In the turn, the NSFC define a set of fundamental (both gauge and spinor)  fields for which the corresponding vacuum equations are identically satisfied~\cite{IJGMMP}. Singularities of these fields (point-, string- or membrane-like) can be treated as particlelike formations participating in a self-consistent collective dynamics~\cite{Khasanov}. We obtain thus a non-trivial set of interacting fields and their singular sources whose properties are determined solely by the $\mathbb B$-differentiability conditions; this approach has been called ``algebrodynamics''~\cite{AD,GR95}. 

However, the complex space-time geometry defined by the structure of $\mathbb B$-algebra embarrass the physical applications of biquaternions and, particularly, the interpretation of results obtained in the framework of the algebrodynamical approach. In our previous works~\cite{GR05,PIRT11} a number of constructions has been presented which, rather naturally, relate complex pre-geometry with real geometry of the Minkowski space or its extensions. These constructions based on $\mathbb B$-invariant {\it bilinear maps} lead to a peculiar kinematics of the particlelike formations; they are described in Section 2. 

In Section 3 we impose some additional restrictions on the particles' kinematics motivated by the properties of the primary $\mathbb B$-field and its twistor partner. These properties define the transport of twistor field and, in addition, the transport of singulartities, caustic-like structures. In result, one obtains a remarkable classification of the properties of singularities-particles themselves. Some concluding remarks are made in the last Section 4. 

%In Section 4, we present and discuss the formerly obtained fundamental relation between the two Lorentz invariants, Minkowski interval and geometric phase, which naturally arise from the primary structure of complex geometry. This relation allows for classical geometrical explanation of the quantum interference phenomena and, in particular, results in a remarkable relativistic generalization of the de Broglie's condition. The latter turns out to be closely related to the Feynmann  representation of the wave functions and to the Bohr-Sommerfeld quantization condition. Some concluding remarks are made in the last section 5. 

\section{Real geometry determined by the algebra of biquaternions}
Biquaternion algebra $\mathbb B$ is isomorphic to the full $2\times 2$ matrix algebra over $\mathbb C$, $Mat(2,\mathbb C)$, so that for any $Z\in \mathbb B$ one has, say, the following representation:
\begin{equation}\label{matrix}
Z={\bf I}z_0 + {\bf \Sigma} \cdot {\bf z} 
=\left(
\begin{array}{cc} 
z_0 + z_3 & z_1 - \imath z_2 \\
z_1 + \imath z_2 & z_0 - z_3
\end{array}
\right),
\end{equation}
where $\{z_0,\bf z\} \in \mathbb C$, {\bf I} is the unit matrix and ${\bf \Sigma} =\{\sigma_a\}, a=1,2,3$ is the triplet of Pauli matrices.  

Consider now a Hermitian ($M=M^+$) matrix, 
\begin{equation}\label{M-element}
M:= Z Z^+ =\left( 
\begin{array}{cc}
T +R_3 & R_1 - \imath R_2 \\
R_1 + \imath R_2  & T - R_3 
\end{array}
\right) = T + \bf \Sigma \cdot \bf R, 
\end{equation}
in which quantities $\{T,\bf R\}$ depend on the complex coordinates and their conjugates as follows:
\begin{equation}\label{realcoord}
\begin{array}{l} 
T:= z_0 \bar z_0 + \bf z \cdot \bf \bar z, ~~
{\bf R}: ={\bf P} +{\bf Q}, \\
{\bf P}=z_0 \bf \bar z + \bar z_0 \bf z,~~
{\bf Q}=\imath~ \bf z \times \bf \bar z
\end{array}
\end{equation}
Bilinear in $Z$ quantities $\{T,\bf R\}$ could represent the real time and space coordinates, respectively; however, more justified interpretation will be discussed below. 

One observes that  spatial coordinates are defined as the sum of two different parts, and the two 3D subspaces $\bf P$ and $\bf Q$ are easily seen to be orthogonal, 
\begin{equation}\label{ortho}
{\bf P} \cdot {\bf Q} = 0.
\end{equation}
Making now use of the well-known properties of determinants, $\det(Z Z^+) = (\det Z)(\det Z^+) =\vert \det Z \vert^2$, one obtains the following remarkable identity~\cite{PIRT11}:
 \begin{equation}\label{quadric}
 T^2 - {\bf R}^2 =S^2, ~~~S:=\vert \det Z \vert = \vert z_0^2 - {\bf z}^2 \vert,
\end{equation}
which, in account of (\ref{realcoord}) and the orthogonality relation (\ref{ortho}), takes the form
\begin{equation}\label{PQquadric}
T^2 -{\bf P}^2-{\bf Q}^2=S^2.
\end{equation}
It is worthy to note that, since $S^2 = \Re(S)^2 + \Im(S^2)$, equation (\ref{PQquadric}) represents a remarkable {\it nine squares identity} closely related to the structure of $\mathbb B$-algebra. Of course, identity (\ref{PQquadric}) can be immediately written out in an explicit coordinate form. 
%?? Brioski Rybakov from octonions the same ??
 
Consider now  a particular case when the component $z_0$ is zero. Then equation (\ref{quadric}) reduces to the {\it identity of six squares} for $\bf z$ of the form
\begin{equation}\label{quadreduc}
\begin{array}l
t^2 - {\bf q}^2 =s^2, \\
t= {\bf z}\cdot \bar {\bf z} , ~~ q= \imath~ {\bf z} \times \bar {\bf z}, ~~s=\vert {\bf z}^2 \vert,
\end{array}
\end{equation}
which has been previously obtained and discussed in ~\cite{GR05}. Importantly, the identity  (\ref{quadreduc}) remain valid, together with (\ref{quadric}), if even $z_0 \ne 0$.

Under transformations from the group $SO(3,\mathbb C)$ which preserve the multiplication law of $\mathbb B$-algebra, both ${\bf z}^2$ and $z_0$ are evidently invariant, so that the quantity $S$ (and the reduced one $s$) can be identified with the {\it Minkowski interval}. Under {\it real} 3D-rotations, the quantities $T$ and $\bf R$ evidently form a scalar and vector, respectively. On the other hand, when the angle of rotation is imaginary, these quantities transform in a way different from the Lorentz transformations. For a reduced case such transformations have been studied in~\cite{GR05}. 

However, if we consider instead the {\it left shift} transformation $Z \mapsto AZ,~ A\in SL(2,\mathbb C)$, then the corresponding Hermitian matrix $M=ZZ^+$ transforms according to the canonical law, $M \mapsto A M A^+$, so that the quantities $\{T,\bf R\}$ undergo standard Lorentz transformations. Note that the {\it dual} mapping $Z\mapsto Z^+Z$ and related {\it right shift} transformation $Z \mapsto ZA$ define in fact the same Lorentz 4-vector $\{T,\bf R\}$ but for the point in $\mathbb {C\bf M}^3$ corresponding to  complex conjugate coordinates $\bf z \rightarrow \bar z$. 
    
One should, however, bear in mind that the introduced quantities $\{T,\bf R\}$, being bilinear in $Z$, are not {\it translationary invariant}. That is,  for two points $Z_1$ and $Z_2$ and corresponding real coordinates $T_1,T_2$ and ${\bf R}_1, {\bf R}_2$ the {\it holonomy} conditions  $\Delta T = T_2-T_1,~\Delta {\bf R} = {\bf R}_2 - {\bf R_1}$ are no longer valid. Here $\Delta T$ and $\Delta {\bf R}$ are the quantities 
expressed through the differences of complex coordinates $\Delta Z$ and their conjugates $\Delta Z^+$ making use of the same definitions (\ref{realcoord}). 

In view of this fact, one should identify the 4-vector $\{T,\bf R\}$ with some other quantity attributed to a point of $\mathbb C{\bf M}^4$, precisely with a set of characteristics of a {\it material pointlike source}. We shall return to this problem below. As for the coordinates themselves, one can define only their increments $\{\delta T, \delta \bf R\}$ using corresponding definitions (\ref{realcoord}), that is:
\begin{equation}\label{increments}
\begin{array}l
\delta T:= dz_0 d\bar z_0 + d{\bf z}\cdot d \bar {\bf  z},~~ \delta {\bf R}:=\delta {\bf P} + \delta {\bf Q},\\
\delta {\bf P}:= dz_0d\bar {\bf z} +d\bar z_0 d {\bf z},~~\delta {\bf Q}:=\imath~\delta {\bf z}\times \delta \bar {\bf z} \\ 
\delta S:=\vert dz_0^2 - d{\bf z}^2\vert.
\end{array}  
\end{equation}
However, these increments do not constitute full differentials, so that corresponding real coordinates turn our to be {\it unholonomic}. This property, though seeming rather exotic, leads up to a number of interesting concepts and conclusions (``random time'', geometric origin of time irreversibility, effective curvature, etc.) some of which have been discussed in~\cite{YadPhys}. 

As the quantities $\{T,\bf R\}$ themselves, the increments of real coordinates satisfy the identities similar to (\ref{quadric}) and (\ref{PQquadric}), that is
\begin{equation}\label{incrquadr}
\delta T^2 - \delta {\bf R}^2 =\delta S^2, 
 \end{equation} 
\begin{equation}\label{incrPQquadr}
\delta T^2 - \delta {\bf P}^2 -\delta {\bf Q}^2 =\delta S^2, 
 \end{equation} 
as well as the orthogonality condition  
\begin{equation}\label{incrortho}
\delta {\bf P}\cdot \delta {\bf Q} =0.
\end{equation}
Note that infinitesimal Minkowski interval $\delta S$ in (\ref{increments}) is necessarily {\it non-negative}, so that one deals here with only the {\it causal} part of the Minkowski-like space-time. On the other hand, the time increment $\delta T$, according to the definition (\ref{increments}), is positive definite, $\delta T >0$, so that one observes here a sort of {\it time irreversibility}: any displacement in the primary complex space results in an increase of time in the induced real space.       

Consider now a pointlike source moving in the primary complex space $\mathbb C {\bf M}$. Then, for the case of nonzero interval, one can define a real 4-velocity vector ($\mu=0,1,2,3$)
\begin{equation}\label{4velocity}
V^\mu:=\frac{\delta X^\mu}{\delta S},~\delta X^\mu:=\{\delta T,\delta {\bf R}\}, 
\end{equation}
which, in account of (\ref{incrquadr}), is identically unit, 
\begin{equation}\label{unit}
V_\mu V^\mu =\left(\frac{\delta T}{\delta S}\right)^2 - \left(\frac{\delta {\bf R}}{\delta S}\right)^2 =1, 
\end{equation}
while the 3-velocity ${\bf V}:=\frac{\delta {\bf R}}{\delta T}$ is always subluminar, 
\begin{equation}\label{3velocity}
\delta T (1-V^2)=\delta S >0, \rightarrow V^2 <1.
\end{equation}

 Note finally that, in view of the identity (\ref{incrPQquadr}) and orthogonality condition (\ref{incrortho}) the full 3-velocity is represented by ``partial'' velocities
 \begin{equation}\label{velocdef}
{\bf u}:=\frac{\delta {\bf P}}{\delta T}, ~~{\bf w}:=\frac{\delta {\bf Q}}{\delta T}
 \end{equation}
corresponding to the motions in the two real 3-subspaces $\bf P$ and $\bf Q$ and mutually orthogonal,
 \begin{equation}\label{partvel}
 V^2 = u^2+w^2,~~~ {\bf u} \cdot {\bf w} =0. 
 \end{equation}       

 In the second case of zero interval, $\delta S=0$, when the complex cone  $\det Z =\delta z_0^2 -\delta {\bf z}^2 =0$ is mapped into the future part of the real light cone $\delta T^2 - \delta {\bf R}^2 =0,~~\delta T >0$, the full 3-velocity is equal to the speed of light, $V=1$, and for partial velocities we have instead of (\ref{partvel}):
\begin{equation}\label{urus1}
u^2+w^2 =1, ~~~{\bf u}\cdot {\bf w} =0.
\end{equation} 
Remarkably, in this case our $\mathbb B$-geometry leads up to the geometry proposed earlier by I. Urusovskii~\cite{Urus3,Urus1,Urus2} in the framework of his ``3+3 treatment of the Special Relativity''. From his viewpoint, the speed of light is not only invariant but {\it universal}, i.e. all material points have one and the same velocity in the full 6D space.

We have seen thus that the $\mathbb B$-induced real geometry not only reproduces the principal properties of the Minkowski space-time but impose, in addition, some restrictions on admissible kinematics of pointlike particles. 
In the next section we shall consider other kinematical relations which arise on the base of $\mathbb B$-analysis and corresponding twistor structures.  

\section{$\mathbb B$-analysis, twistors and kinematics in $\mathbb B$-induced real space} 
In the {\it algebrodynamical program}~\cite{AD,GR95,Qanalisys,YadPhys} primary physical fields are $\mathbb B$-differentiable functions. Conditions of $\mathbb B$-differentiability, a non-commutative generalization of the Cauchy-Riemann conditions from complex analysis, are considered as the primary field equations; in the principal case they reduce to the  overdetermined matrix system of equations for differentials,  
\begin{equation}\label{GSE}
d\xi = \Phi dZ \xi
\end{equation}
 for a 2-spinor $\xi=\{\xi_A (Z)\}$ and complex 4-vector $\Phi=\{\Phi_{AB} (Z)\}$ fields ($A,B=0,1$, and $Z$ is the  $\mathbb B$-coordinate matrix). The structure of general solution of (\ref{GSE}) can be easily obtained by the following its transformation
\begin{equation}\label{Twist}
d\xi=\Phi d (Z\xi) - (\Phi dZ) \xi, \rightarrow \xi = \xi(\tau),
\end{equation}
 where $\tau:=Z \xi$ together with $\xi$ constitute the (projective) {\it twistor} of complex space $\mathbb C{\bf M}$. Therefore, in an invariant form general solution of (\ref{GSE}) is represented by a system of two algebraic equations~\cite{IJGMMP}
\begin{equation}\label{general}
\Pi^{(C)}(\xi,Z\xi)=0,~~C=1,2, 
 \end{equation}
 with $\Pi^{(C)}$ being two arbitrary analytical functions of four twistor variables. Making use of the projective invariant structure of the incidence relation $\tau=Z\xi$, one can reduce (\ref{general}) to a simpler form of one equation with one {\it homogeneous} generating twistor function $\Pi$,    
\begin{equation}\label{homogen}
\Pi(\xi,Z\xi)=0,
\end{equation}
from which only the {\it ratio} $G$ of two spinor components $\xi_A$ can be determined at each fixed point of $\mathbb C{\bf M}$. 

On the Minkowski subspace $\bf M$ of ${\mathbb C} \bf M$ represented by Hermitian matrices of coordinates $Z\mapsto X=X^+$, solution of (\ref{GSE}) in the form (\ref{homogen}) reproduces the celebrated {\it Kerr-Penrose} theorem~\cite{Kerr,Penrose} completely describing the structure of {\it null shear-free congruences} (NSFC) on $\bf M$ mentioned in Section 1. Thus, $\mathbb B$-differentiable functions explicitly give rise to a complex extension of the Minkowski twistors and, geometrically, to a complex extention of NSFC. 

{\it Caustics} of the rays of NSFC corresponding to multiple roots of the equation (\ref{homogen}) and  defined, therefore, by the condition 
\begin{equation}\label{caust}
\frac{d\Pi}{dG}=0
\end{equation}
Structure and dynamics of caustics correspond to those of singularities of fundamental fields defined by NSFC which can be treated as particlelike formations. For more details we refer the reader, say, to~\cite{Sing, IJGMMP} and references therein.
 
Let us consider now the (infinitesimal) transformations of complex coordinates which leave invariant the primary twistor field and the caustic structure. Specifically, from the equation (\ref{homogen})  as the {\it condition of twistor preservation} one obtains $(dZ) \xi=0$, which in components leads up to the following two real relations for differentials $dZ$: 
\begin{equation}\label{twistpreserve}
dy_0 = {\bf n}\cdot d{\bf y},~~~ 
d{\bf x} = {\bf n}~dx_0 + {\bf n} \times d{\bf y},
\end{equation}
where $dx_\mu$ and $dy_\mu$ are the real and imaginary parts of the increments of four complex coordinates $dz_\mu =dx_\mu + \imath dy_\mu$. The  (projective invariant) {\it unit director 3-vector} $\bf n$  is formed from the components of the principal spinor $\xi$ as follows:
\begin{equation}\label{director}
{\bf n}:=\frac{\xi^+{\bf \Sigma} \xi}{\xi^+\xi},~~{\bf n}^2\equiv 1.
\end{equation}   
Of course, conditions  (\ref{twistpreserve}) can be easily inverted, 
\begin{equation}\label{twistpreserveINV}
dx_0 = {\bf n}\cdot d{\bf x},~~~ 
d{\bf y} = {\bf n}~dy_0 - {\bf n} \times d{\bf x},
\end{equation} 
 and the increments identically satisfy the {\it complex null cone} condition:
  \begin{equation}\label{nullcone}
 dz_0^2 - d{\bf z}^2 =0. 
 \end{equation}
 % IN MATRIX FORM $dX=NdY$???   

It is worthy to note that under transition to the Minkowski space-time, when  $Z\mapsto X=X^+$, conditions (\ref{twistpreserve}) reduce to 
$d{\bf x}={\bf n} dt$, where $dt:=dx_0$ is the real Minkowski time. The latter condition corresponds to the well-known transport of twistor field along a 3-direction (defined by the initial value of the principal spinor) with velocity of light. 
  
 The complex cone condition can be equivalently written out in the form of two real conditions:
 \begin{equation}\label{twistgeom}
 \begin{array}l
 dx_0^2 + d{\bf y}^2 = dy_0^2 +d{\bf x}^2 =\frac{1}{2}\delta T,\\
 d{\bf x}\cdot d{\bf y} =dx_0 dy_0.
 \end{array}
 \end{equation}
 The last equality in the first equation follows from comparison of the latter with the previous definition of the time increment in (\ref{increments}). Note that the topology of the twistor preserving 6D increment space (actualy, that of the complex null cone (\ref{nullcone})), in a consequence of (\ref{twistgeom}), is $S^3\times S^2 \times R_+$). 
 
From three equations (\ref{twistgeom}) one can, in particular, express the invariant $\vert dz_0\vert^2=dx_0^2+dy_0^2$ through the invariants of spacial increments, 
 \begin{equation}\label{traceincrement}
 dx_0^2+dy_0^2 = \sqrt{(d{\bf x}^2 -d{\bf y}^2)^2 +(2d{\bf x}\cdot d{\bf y})^2},
 \end{equation} 
or, after a simple transformation, obtain another important relation:
\begin{equation}\label{Mink2}
(\delta {\bf x}^2 +\delta {\bf y}^2)^2 - \vert 2 d{\bf x}\times d{\bf y}\vert^2 = (dx_0^2+dy_0^2)^2. 
\end{equation}
Thus, despite the fact that the previously defined complex and real Minkowski intervals in the twistor preserving geometry are both zero, 
one can, in account of (\ref{Mink2}), consider as infinitesimal Minkowski interval $ds $ the invariant
\begin{equation}\label{redefinterval}
 ds:=\vert dz_0\vert^2 =dx_0^2+dy_0^2 \ge 0,
 \end{equation}
 while for the redefined time $\delta t$ and space $\delta \bf r$  increments of the coordinates take the quantities
\begin{equation}\label{redefincrmnts}
\delta t:=\delta {\bf x}^2 +\delta {\bf y}^2 \ge 0,~~\delta {\bf r}: = 2 d{\bf x}\times d{\bf y}.
 \end{equation}     
By this identifications, the principal relation (\ref{Mink2}) of the twistor preserving geometry takes the Minkowski-like  form (again with non-negatively definite increments $\delta t$ and $\delta s$):
\begin{equation}\label{Mink2redef}
\delta t^2 -\vert \delta {\bf r} \vert^2 = \delta s^2 
 \end{equation}
 (compare with the reduced identity (\ref{quadreduc}) for corresponding increments in account of the complex null cone condition $dz_0^2 =d{\bf z}^2$). 
 
On the other hand, for the full increments of the coordinates (\ref{increments}) one obtains:
\begin{equation}\label{incrtwist}
\begin{array}l 
\delta T =\sqrt{(d{\bf x}^2 -d{\bf y}^2)^2 +(2d{\bf x}\cdot d{\bf y})^2}+d{\bf x}^2+d{\bf y}^2,\\
\delta {\bf P} = 2 ({\bf n}\cdot d{\bf x}) d{\bf x} +({\bf n}\cdot d{\bf y}) d{\bf y}, ~~\delta {\bf Q} = 2 d{\bf x} \times d{\bf y},
\end{array} 
\end{equation}
and the relations $\delta T^2 - \delta {\bf P}^2 - \delta {\bf Q}^2  =0, ~\delta {\bf P} \cdot \delta {\bf Q} =0$ are valid for any $d{\bf x},d{\bf y}$. 

%UNDER AUTOMORPHISMS (see page 7).

%MICROSCOPIC TIME SCALE (also page 7)

 %PARTICULAR CASES (see page 2)

\section{Caustic preserving geometry and effective kinematics in real space}
     
Let us now impose, in addition to the twistor preserving condition, the analogous requirement on the caustic structure. Specifically, let us consider the increments of complex coordinates which transfer any caustic point  into  another one. From the defining equation (\ref{caust}) of the caustic locus, assuming the constancy of twistor field, one obtains the preservation condition for the matrix components:
\begin{equation}\label{caustpreserv}
dZ_A^B=\Pi_A \xi^B d\lambda, ~~A,B=0,1,
\end{equation}
where $\Pi_A=d\Pi / d\tau^A$ are the derivatives of generating function $\Pi$ w.r.t. two of its twistor arguments, and $d \lambda$ is one sole infinitesimal complex parameter.  

From (\ref{caustpreserv}) it follows that, since the twistor-valued $\Pi_A$ and $\xi_A$ remain constant, the combined twistor-caustic preserving transport takes place along a {\it complex null straight line}. If $dz_0 \ne 0$, condition (\ref{caustpreserv}) can be equivalently rewritten in a vector form, 
\begin{equation}\label{transfer1}
d{\bf z} = {\bf m}~dz_0, ~~{\bf m}^2 =1, 
\end{equation}
where ${\bf m}$ is a unit complex 3-vector, 
\begin{equation}\label{u-vector}
{\bf m}: ={\bf a} + \imath {\bf b}; ~\rightarrow {\bf a}^2 - {\bf b}^2 =1,~ {\bf a}\cdot {\bf b}=0.
\end{equation}
In the other case, when $dz_0=0$, condition (\ref{caustpreserv}) gives:
\begin{equation}\label{transfer2}
d{\bf z} = {\bf l}~d\tau, ~~{\bf l}^2 =0, 
\end{equation}  
 with a null complex 3-vector,     
 \begin{equation}\label{n-vector}    
 {\bf l}={\bf c}+\imath {\bf d}; \rightarrow {\bf c}^2 = {\bf d}^2,~~{\bf c}\cdot {\bf d} =0, 
 \end{equation} 
 and an infinitesimal complex parameter $d\tau$.
 
 Calculating then the increments of real temporal and spacial coordinates (\ref{increments}), in the first case
 (\ref{transfer1}), one obtains
 \begin{equation}\label{Rtransfer1}
 \delta T = {\bf a}^2~ ds, ~~\delta {\bf P} =  {\bf a}~ ds, ~~\delta {\bf Q} =  ({\bf a} \times {\bf b}) ds, 
 \end{equation}
 while the increment of proper time parameter {\it turns to be null}, $\delta S =0$! The increment of real parameter $ds$ is defined by the square of modulus  of the zeroth complex coordinate, $ds=2 \vert dz_0\vert^2$. Thus, in account of (\ref{incrPQquadr}), in the induced real space-time singularity moves with the full velocity equal to the speed of light. The latter consists of the two components,   
${\bf u} = \delta {\bf P}/\delta T < 1$ and ${\bf w} = \delta {\bf Q} / \delta T <1$ which both are smaller than $c$ and mutually orthogonal. 

In the second case (\ref{transfer1}), when $dz_0 =0$, one has, respectively:
\begin{equation}\label{Rtransfer2}
 \delta T = {\bf c}^2~ ds, ~~\delta {\bf P} \equiv 0, ~~\delta {\bf Q} =  ({\bf c} \times {\bf d}) ds, 
 \end{equation}
 where $ds=2 \vert d\tau \vert^2$, and again the increment of proper time parameter $\delta S =0$. One observes that singularity again moves with the the speed of light but in this case only one component of velocity is nonzero. 

We have already mentioned that the quantities $T,\bf R$ as well as their increments $\delta T, \delta \bf R$ are not invariant under translations and unholonomic. Therefore, these (bilinear in complex coordinates $z_0, \bf z$) quantities should be interpreted not as coordinates but rather as {\bf energy} and {\bf momentum} of a corresponding particlelike formation.

\section{Conclusion}
Existence of complex pre-geometry seems to be unavoidable in algebrodynamical approach. Indeed, in the framework of the latter, the whole physical geometry and dynamics should be derived solely from the properties of a space-time algebra. However, the only candidate for this role is the algebra of biquaternions with its $SO(3,\mathbb C) \sim SL(2,\mathbb C$)  automorphism group and 4D complex vector space. 

Thus, to construct the unified biquaternionic algebrodynamics, one should discover the links between the complex geometry $\mathbb C\bf M$ and real Minkowski space-time $\bf M$.  In the most natural way one can do it via bilinear mapping $Z \mapsto X:=ZZ^+$ used in the paper. Then the induced kinematics of particles-singularities turns to be Lorentz invariant. However, quantities $X$ are not translation  invariant and correspond rather to the energy-momentum 4-vector of a particle. As for the real coordinates themselves, they can be identified with real parts of complex coordinates while imaginary parts constitute the 4D timelike evolution parameter, see (\ref{twistpreserve}).  Note that multidimensional time can be the origin of real time irreversibility, of quantum uncertainty etc. \cite{YadPhys}. 

In the paper, we have also studied the effective kinematics of a singular point defined by the requirements of preservation of the primary twistor field and the singularity itself. These conditions arise in the framework of $\mathbb B$-analysis and lead to two different types of the transport of a singular point along a straight line in both of which, however, the full real space velocity of transport is equal to the speed of light. Note, however, that for the points on a {\it focal line} (unique Worldline) \cite{Khasanov} the effective kinematics (dynamics) can be much more complicated since the primary spinor field of the congruence $G$ is indefinite on the worldline itself. Complete investigation of this case will be presented elsewhere.

%\section*{References}

\end{document}